\documentclass{article}
\usepackage[T1]{fontenc}
\usepackage[utf8]{inputenc}
\usepackage{ismir} 
\usepackage{amsmath,cite,url}
\usepackage{graphicx}
\usepackage{color}
\usepackage{mathtools}
\usepackage{amsfonts}
\usepackage{multirow}
\usepackage{booktabs}
\usepackage{pifont}

\newcommand{\V}[1]{\mathbf{#1}}

\title{Improving BERT for symbolic music understanding using token denoising and pianoroll prediction}

\twoauthors
  {Jun-You Wang} {Institute of Information Science \\ Academia Sinica}
  {Li Su} {Institute of Information Science \\ Academia Sinica}

\def\authorname{J.-Y. Wang and L. Su}

\begin{document}

\maketitle

\begin{abstract}
We propose a pre-trained BERT-like model for symbolic music understanding that achieves competitive performance across a wide range of downstream tasks. To achieve this target, we design two novel pre-training objectives, namely token correction and pianoroll prediction. First, we sample a portion of note tokens and corrupt them with a limited amount of noise, and then train the model to denoise the corrupted tokens; second, we also train the model to predict bar-level and local pianoroll-derived representations from the corrupted note tokens. We argue that these objectives guide the model to better learn specific musical knowledge such as pitch intervals. For evaluation, we propose a benchmark that incorporates 12 downstream tasks ranging from chord estimation to symbolic genre classification. Results confirm the effectiveness of the proposed pre-training objectives on downstream tasks.
\end{abstract}

\section{Introduction}\label{sec:introduction}
In recent years, music information retrieval (MIR) research in the symbolic music domain has undergone a paradigm shift from the development of task-specific models to the adoption of the pre-training/fine-tuning paradigm, where a model is first pre-trained on a large-scale dataset with self-supervised learning (SSL) objectives and then fine-tuned on specific downstream tasks~\cite{Chou2021midibert, Liang2024pianobart, Zhao2024adversarial, Shen2023more, Zeng2021musicbert, Wang2021musebert, Sailor2024rnbert}. One of the advantages is that this paradigm is adaptable to downstream tasks with only minimum task-specific designs. Inspired by the advance of natural language processing (NLP)~\cite{openai2023gpt4, Touvron2023llama, Beeching2023open}, pre-trained symbolic music understanding models like MidiBERT~\cite{Chou2021midibert} and MusicBERT~\cite{Zeng2021musicbert} typically adopt a similar architecture and pre-training objective as BERT, i.e., masked language modeling (MLM) in text sequences~\cite{Devlin2019bert}. Discussions for improvement have been therefore focused on the masking strategy during pre-training~\cite{Zhao2024adversarial, Shen2023more}, leading to improved performance in downstream tasks.

However, currently there are several issues that have to be overcome. 
First, there is a limitation of using only MLM in symbolic music modeling due to the fundamental differences between music and text.
Unlike words, symbolic music events are usually the notes of a specific music scale performed at specific time intervals.
The MLM objective does not implicitly encode such music knowledge, but treats all note tokens as individuals, regardless of their possible tonal or metrical implications. 
Modifications regarding pre-training objectives should be made to help the model learn such aspects of symbolic music better. Second, there is a lack of a comprehensive evaluation for pre-trained models for symbolic music. Previous works are only evaluated for five or fewer downstream tasks quantitatively~\cite{Chou2021midibert, Liang2024pianobart, Zeng2021musicbert, Zhao2024adversarial, Shen2023more, Wang2021musebert}, which could be improved to reveal more advantages and disadvantages of a pre-trained model.

To address these issues, we make two main contributions in this work. First, we propose two novel self-supervised learning objectives for pre-training. The first one, namely the \emph{token denoising} objective, is a modified version of the MLM objective. Instead of randomly replacing note tokens with a pre-defined \texttt{[MASK]} token, we randomly corrupt the attributes of note tokens by adding small random noises and train the model to reconstruct the original tokens. Such a token de-noising objective implicitly encodes the actual distance information of note attribute values. The second one, namely the \emph{pianoroll prediction} objective, has the model infer pitch and chroma distribution from the corrupted input note sequence, which resembles the prediction of pianoroll representation. This helps the model implicitly learn the idea of pianoroll, which encompasses important features such as the temporal information and the interval between notes, which are crucial in downstream tasks such as melody extraction~\cite{Uitdenbogerd1999melodic, Simonetta2019convolutional, Kosta2022deep}, roman numeral analysis~\cite{Lopez2021augmentednet, Sailor2024rnbert, Karystinaios2023roman}, etc. As a result, the pre-trained model acquires domain knowledge on symbolic music and performs better on downstream tasks.

Second, to verify the effectiveness of pre-trained models, we conduct a comprehensive evaluation by adapting the models to a total of 12 downstream classification tasks, which far exceeds the scale of evaluation in previous works~\cite{Chou2021midibert, Liang2024pianobart, Zhao2024adversarial, Shen2023more, Zeng2021musicbert, Wang2021musebert}. To achieve this goal, we propose a unified protocol for fine-tuning, and recast various symbolic music understanding tasks with publicly available datasets into this framework.
By fine-tuning a pre-trained model on such a wide range of downstream tasks, we do not only demonstrate the effectiveness of the proposed methods, but also pave the way for the evaluation of future work. As a preliminary step toward formalizing the evaluation of pre-trained models in symbolic music, we invite the research community to include more diverse downstream tasks and further refine the protocol to facilitate the future development of different forms of symbolic pre-trained models.

\section{Related work}\label{sec:related_work}
Symbolic music data is an abstract representation of music~\cite{Le2024natural}, which represents a musical piece with a sequence of notes. Each note contains at least three attributes: onset timing, note duration, and note pitch. Additional attributes such as tempo, time signature, staff information, etc., may also present. 
Symbolic music understanding aims to infer high-level information from it, such as melody~\cite{Uitdenbogerd1999melodic, Kosta2022deep}, motifs~\cite{Hsiao2023bps}, repeated patterns~\cite{collins2013discovery, meredith2002algorithms}, functional harmony~\cite{Lopez2021augmentednet, chen2018functional}, texture~\cite{Lin2024s3}, genre~\cite{Schreiber2015improving}, structural boundary~\cite{Olivan2023symbolic}, and emotion~\cite{Hung2021emopia, Zhao2023multimodal}. 


As symbolic music understanding tasks highly involve the domain knowledge of music, annotating ground-truths for these tasks typically have to be done by experts, sometimes assisted with computational tools~\cite{Foscarin2020asap}. As a result, available data for symbolic music understanding tasks are usually scarce. To address this issue, previous work has proposed to generate training data automatically~\cite{Zhang2022atepp} and obtaining annotation from the Internet~\cite{Ycart2018amaps}. Recently, there have been attempts to directly utilize large language models (LLMs) for symbolic music understanding tasks~\cite{Zhou2024can, Yuan2024chatmusician}, which removes the demand of a large-scale dataset; however, the results are unsatisfactory. In tasks related to music reasoning such as key estimation and harmony analysis, the best LLM only achieves an accuracy slightly better than random guess, possibly due to the domain gap between text and symbolic music~\cite{Zhou2024can}.

Another research direction is to adapt a pre-trained model in the symbolic music domain to downstream tasks.
In the past few years, the surge of pre-trained models that can be easily adapted to various downstream tasks has led to a paradigm change in multiple domains~\cite{Bommasani2021opportunities, Won2024foundation}, such as text~\cite{Devlin2019bert, Touvron2023llama, openai2023gpt4}, image~\cite{Radford2021learning}, speech~\cite{Yang2021superb, Chen2022wavlm, Huang2024dynamic}, and music in the audio domain~\cite{Won2024foundation, li2024mert, Liao2024music, Castellon2021codified}. In symbolic music domain, there are also efforts to build pre-trained models for symbolic music understanding~\cite{Chou2021midibert, Zhao2024adversarial, Shen2023more, Zeng2021musicbert} and symbolic music generation~\cite{Liang2024pianobart, Wang2021musebert, Wu2024melodyt5}. Due to the similarity between text and symbolic music~\cite{Le2024natural}, previous pre-trained models on symbolic music understanding mostly employs similar architectures and pre-training objectives as in NLP. For example, \cite{Zeng2021musicbert, Chou2021midibert, Zhao2024adversarial, Shen2023more} are all based on BERT~\cite{Devlin2019bert}.
In this work, we propose novel objectives that take into account the distinct characteristics of symbolic music for pre-training.

\begin{figure*}[t]
    \centering
    \includegraphics[width=\textwidth, trim={0 0.35cm 0 0.15cm},clip]{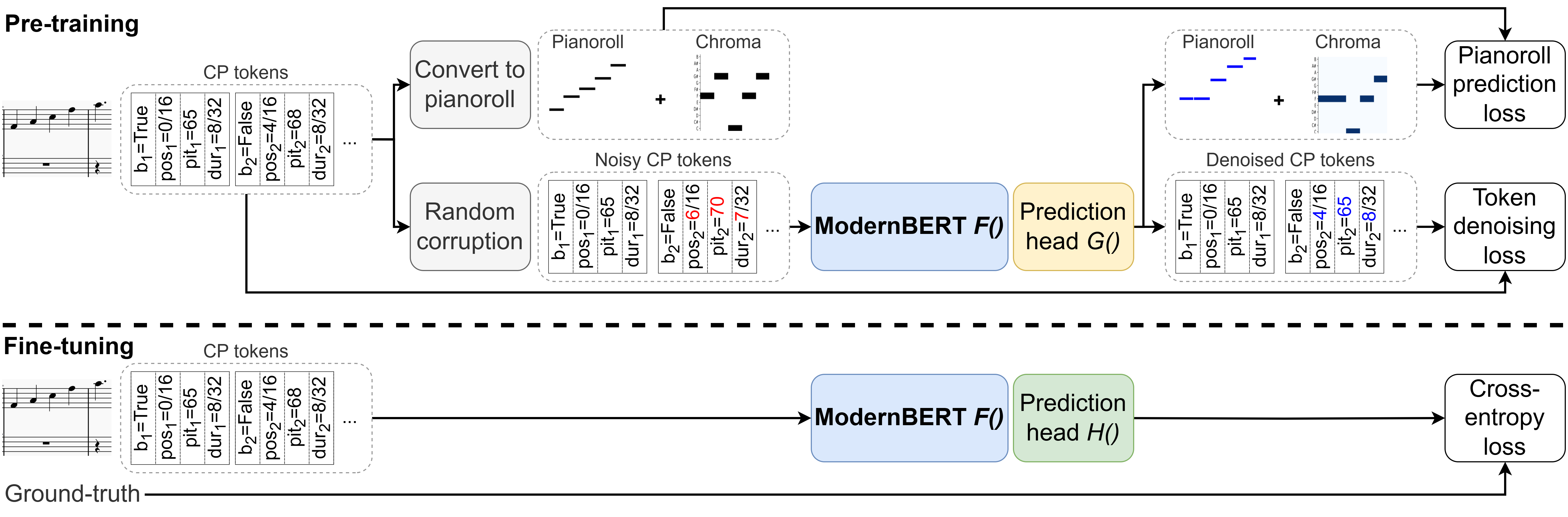}
    \caption{An overview of the proposed method.}
    \label{fig:overview}
\end{figure*}

\section{Proposed method}\label{sec:proposed_method}
Figure~\ref{fig:overview} gives an overview of the proposed method. The input symbolic music representation is a music piece $\V{X}$ composed of $N$ notes, i.e., $\V{X} \coloneqq \{\V{x}_n\}_{n=1}^{N}$, and $\V{x}_n:=(\mathrm{o}_n, \mathrm{p_n}, \mathrm{d_n})$ represents the $n$-th note with its onset, pitch, and duration in $\V{X}$, and a list of downbeat timings $\V{DB}$ composed of $M$ downbeat timings, i.e., $\V{DB} \coloneqq \{\mathrm{db}_m\}_{m=1}^{M}$. For MIDI data (including performance MIDI), an unreliable yet doable way to obtain the raw $\V{DB}$ information is to retrieve the tick information (e.g., the number of ticks per beat and downbeat) in MIDI. The proposed pre-trained model $F(\cdot)$ generates a contextualized representation at the note level. During pre-training, we add a prediction head $G$ on top of $F$ and train $G(F(\cdot))$ with unlabeled data through self-supervised learning objectives.
Then, during fine-tuning, we remove $G$ and add another prediction head $H(\cdot)$ on top of $F$, and train both $F$ and $H$ in a downstream classification task.
We aim to improve $F$ so that $H(F(\cdot))$ could achieve better overall performance in a wide range of downstream tasks.

\subsection{Tokenization and segmentation}\label{subsec:tokenization} 
Following MidiBERT~\cite{Chou2021midibert}, we adopt a modified version of the Compound Word (CP) representation~\cite{Hsiao2021compound} for tokenization. Each note $\V{x}_n$ is represented by a CP token with four attributes, $(\textrm{b}_n, \textrm{pos}_n, \textrm{pit}_n, \textrm{dur}_n)$. $\textrm{b}_n \in [0, 1]$ is a binary flag that denotes whether a note is at the start of a bar; $\textrm{pos}_n \in [0, 15]$ denotes the onset position of a note within a bar rounded to an integer number of 1/4 crotchet beats; $\textrm{pit}_n \in [22, 107]$ denotes note pitch; $\textrm{dur}_n \in [1, 64]$ denotes note duration rounded to an integer number of 1/8 crotchet beats. In practice, we follow MidiBERT by assuming that the duration of a bar is four crotchet beats (i.e., 4/4 or 2/2 time signature). For bars that have a different duration (which can be inferred from $\V{DB}$), we rescale them to four crotchet beats. As for segmentation, following~\cite{Zeng2021musicbert}, we adopt a maximum sequence length of 1024 tokens.

\subsection{Backbone model}\label{subsec:backbone_model}
In previous work, $\text{BERT}_\text{BASE}$~\cite{Devlin2019bert}, a bidirectional encoder with 12 Transformer layers~\cite{Vaswani2017attention}, has been widely used as the backbone model~\cite{Chou2021midibert, Wang2021musebert, Zeng2021musicbert, Zhao2024adversarial, Shen2023more}. In this work, we choose ModernBERT~\cite{Warner2024smarter} as our backbone model. ModernBERT is an enhanced version of BERT that incorporates various techniques to improve training efficiency and performance, such as flash attention~\cite{Dao2022flashattention}, local attention, rotary positional encoding mechanisms~\cite{Su2024roformer}, etc. Although the official ModernBERT model contains 22 layers, we reduce them to 12 for a fair comparison with MidiBERT. 
We refer to our proposed ModernBERT-based model as \textbf{M2BERT}, which stands for \textbf{Modern-MidiBERT}.

\subsection{Pre-training objectives}\label{subsec:pretraining_objectives}
In both BERT and MidiBERT~\cite{Devlin2019bert, Chou2021midibert}, the MLM objective is applied for pre-training. They first randomly choose a fraction of input tokens (15\% of tokens in practice). For the chosen tokens, 80\% of them are replaced with a special \texttt{[MASK]} token, 10\% of them are replaced with a randomly sampled token, and the remaining 10\% tokens are kept intact. Then, the model is trained to recover the original tokens from such \emph{masked} inputs.
However, the MLM objective does not fully account for domain-specific knowledge in music. For example, the fact that ``the pitch difference between E4 and C4 is necessarily a major third'' cannot be inferred simply by successfully predicting a masked token. As a result, the model lacks a mechanism to ensure that the learned embeddings of these two pitches reflect the general relationship shared by all major third intervals.

To address this issue, we propose two pre-training objectives. The first is to train the model to correct corrupted tokens rather than masked tokens, and employ a corruption strategy that is aware of the proximity of notes (named as the \emph{token denoising} objective). The second is to have the model predict not only tokens but also pianoroll-like features, which can capture geometrical information of notes that is well encoded in a pianoroll representation but cannot be trivially inferred from a token sequence alone (named as the \emph{pianoroll prediction} objective). The two pre-training objectives are described below.

\subsubsection{Token denoising} 
Similar to MLM, we also randomly sample notes and corrupt them. But instead of replacing notes with the \texttt{[MASK]} token, we corrupt notes by randomly perturbing the value of the tokens. For an input token $(\textrm{b}_n, \textrm{pos}_n, \textrm{pit}_n, \textrm{dur}_n)$, we obtain a corrupted token $(\tilde{\textrm{b}}_n, \tilde{\textrm{pos}}_n, \tilde{\textrm{pit}}_n, \tilde{\textrm{dur}}_n)$ it by:
\begin{align}
    \tilde{\textrm{b}}_n &= \textrm{rand}(0, 2), \\ 
    \tilde{\textrm{pos}}_n &= \textrm{rand}(\textrm{clip}(\textrm{pos}_n - \textrm{r}_\textrm{pos}), \textrm{clip}(\textrm{pos}_n + \textrm{r}_\textrm{pos})), \\
    \tilde{\textrm{pit}}_n &= \textrm{rand}(\textrm{clip}(\textrm{pit}_n - \textrm{r}_\textrm{pit}), \textrm{clip}(\textrm{pit}_n + \textrm{r}_\textrm{pit})), \\
    \tilde{\textrm{dur}}_n &= \textrm{rand}(\textrm{clip}(\textrm{dur}_n - \textrm{r}_\textrm{dur}), \textrm{clip}(\textrm{dur}_n + \textrm{r}_\textrm{dur})),
\end{align}
where the $\text{rand}(x, y)$ function randomly samples an integer from the interval $\left[x,y\right)$;
the $\text{clip}(z)$ function clips $z$ such that $z$ lies in a valid range of token number (e.g., $[0, 15]$ for $\textrm{pos}_n$); 
the hyperparameters $\textrm{r}_\textrm{pos}$, $\textrm{r}_\textrm{pit}$, and $\textrm{r}_\textrm{dur}$ determine the range of corruption. Setting $\textrm{r}_\textrm{pos}, \textrm{r}_\textrm{pit}, \textrm{r}_\textrm{dur} := \infty$ results in the case of random sampling from all possible tokens. 
The proposed model is trained to recover the original note tokens (i.e., \emph{denoise}) from the corrupted ones with the cross-entropy loss, similar to the MLM objective in BERT.

The reasons behind this corruption strategy are two-fold. 
First, as noted in previous NLP work, using the \texttt{[MASK]} token introduces a discrepancy of distribution between pre-training data and fine-tuning data, making the pre-training process allocate model dimensions to exclusively represent the \texttt{[MASK]} tokens~\cite{Meng2024representation}. 
The same work also shows that randomly replacing \texttt{[MASK]} to other tokens is a suboptimal strategy~\cite{Meng2024representation}. Our proposed objective (i.e., randomly corrupting tokens but within a limited range) represents a compromise strategy to reduce such discrepancy, and is easy to operate on symbolic music data. Second, limiting the range of corruption encourages the model to better learn the proximally relative distance between note tokens, which usually carries musical meanings. For example, suppose $\textrm{r}_\textrm{pit}$ is set to 12 semitones, making the model observe a note corrupted to C6. The model is then guided to infer that the correct pitch must be located between C5 and C7 rather than other pitch values far away from C6. This mechanism is desirable, as it guides the model to learn the knowledge related to the pitch interval.

\subsubsection{Pianoroll prediction}
Another set of the proposed pre-training objective is to make the model predict pianoroll and chroma representations. Two prediction mechanisms are imposed:
\begin{itemize}
    \item The \emph{bar-level} prediction mechanism has each $\V{x}_n$ predict the pianoroll (denoted as $\textrm{PR}_n$) and chromagram (denoted as $\textrm{CM}_n$) of the bar in which the onset of $\V{x}_n$ is located.
    
    \item The \emph{local} prediction mechanism has each $\V{x}_n$ predict the pianoroll and chromagram at the onset position of $\V{x}_n$, i.e., $\textrm{PR}_n[\textrm{pos}_n]$ and $\textrm{CM}_n[\textrm{pos}_n]$.  
\end{itemize}

The two mechanisms represent different types of guidance for the model's prediction capability: bar-level prediction ensures that all notes within the same bar predict the same representation, while local prediction allows each note to individually predict the representation corresponding to its specific timing. In practice, we divide one bar into 16 tatums (1/4 crotchet beats per tatum). Given that pitch tokens range from 22 to 107, the dimensions of $\textrm{PR}_n$ and $\textrm{CM}_n$ are (16, 86) and (16, 12), respectively. We employ the L2 loss with the weight for each loss term being equal. Specifically, for each $\V{x}_n$, we have: 
\begin{align}
\mathcal{L}_n &= \text{L2}(\textrm{PR}_n, \hat{\textrm{PR}}_n) + \text{L2}(\textrm{CM}_n, \hat{\textrm{CM}}_n) \nonumber \\ 
&+ \text{L2}(\textrm{PR}_n[\textrm{pos}_n], \hat{\textrm{PR}}_n[\textrm{pos}_n]) \nonumber \\
&+ \text{L2}(\textrm{CM}_n[\textrm{pos}_n], \hat{\textrm{CM}}_n[\textrm{pos}_n])\,,
\end{align}
where $\hat{\textrm{PR}}_n$ and $\hat{\textrm{CM}}_n$ denote the predicted pianoroll and chromagram from $\V{x}_n$ respectively.
The total loss is therefore the sum of the token reconstruction loss and the pianoroll prediction loss over all notes.

\subsection{Prediction heads}

Two types of prediction heads $H$ are utilized to perform the downstream tasks in the fine-tuning stage. The first type is for note-level prediction tasks, where $H$ is simply a linear layer with Softmax activation. Another type is for sequence-level prediction tasks, where $H$ contains an attention-based weighting average layer, a linear layer and Softmax activation. The attention-based weighting average layer uses two linear layers and a Softmax activation to determine the weight for each note, and then computes the weighting average of all notes' embedding to form the sequence-level embedding. All these settings are derived from MidiBERT's protocol~\cite{Chou2021midibert}, where the only differences are that we remove the dropout layer and reduce the number of linear layers for classification to one, which simplifies $H$ and better aligns with BERT's fine-tuning protocol~\cite{Devlin2019bert}. In this work, three downstream tasks (SGC, PS, and ER, see Section \ref{sec:downstream_tasks}  for details) are trained at sequence level, while all other downstream tasks are trained at note level. The $G$ used for pre-training is also constructed with a linear layer with Softmax activation.

During fine-tuning, the backbone model $F$ and the prediction head $H$ are jointly optimized. For each downstream task, we fine-tune the model with a training set, and select the best model checkpoint based on the model performance on a validation set following the designated evaluation metric of that task.

\section{The SMC benchmark}\label{sec:downstream_tasks}
We conduct an evaluation on 12 different downstream tasks for symbolic music classification. We refer to the combination of all these tasks and datasets as the \textbf{SMC benchmark}, which stands for Symbolic Music Classification benchmark. The benchmark is available at \url{https://zenodo.org/records/15681035}.

Similar to~\cite{Devlin2019bert, Chou2021midibert, Zhao2024adversarial, Shen2023more}, we focus on classification tasks, a common problem formulation of symbolic music understanding. Additionally, we aim to integrate these downstream tasks to achieve the following three objectives. 1) \textbf{Reproducibility.} We only select tasks with publicly available datasets that can be freely downloaded without requiring a data access application; 2) \textbf{Diversity.} We include a wide range of datasets while ensuring that each dataset is used for at most two downstream tasks.\footnote{For example, the dataset compiled by AugmentedNet~\cite{Lopez2021augmentednet} is used for functional harmony analysis, which is decomposed into five subtasks. We only include two of them (chord root and local key) as downstream tasks.} 3) \textbf{Comparability.} We prefer to select tasks that have been discussed in previous work and have publicly reported state-of-the-art results for comparison.

The 12 tasks are listed as follows:

\textbf{Symbolic genre classification (SGC).} This task classifies the music genre of symbolic music (in MIDI format). We adopt the \textit{CD2} part of the Tagtraum genre annotations~\cite{Schreiber2015improving} for this task. To construct a balanced dataset, we select the five most frequent genres in the dataset\footnote{The five genres are Country, Electronic, Pop, RnB, and Rock.} and sample 1,150 songs for each. Then, we divide the dataset into training/validation/test set with 500/150/500 songs for each class. We adopt the accuracy metric for this task.

\textbf{Piano performer style classification (PS).} This task classifies the piano performer of a symbolic music from eight candidate pianists. It was introduced in MidiBERT along with the Pianist8 dataset~\cite{Chou2021midibert}. Instead of following the official train/test split, considering the modest scale of the dataset, we divide it into five folds and perform cross-validation. We adopt the accuracy metric for this task.

\textbf{Emotion recognition (ER).} This task classifies the emotion of symbolic music into four quadrants of the valence-arousal space. Following~\cite{Chou2021midibert}, we use the EMOPIA dataset~\cite{Hung2021emopia} and adopt the accuracy metric. We divide the dataset into five folds and perform cross-validation.

\textbf{Beat note prediction (BP)}. This task identifies beat notes from a performance MIDI note sequence, where each note contains only onset, offset, and pitch information, with onset and offset measured in seconds rather than quantized beat numbers.
BP is a subtask of the Performance MIDI-to-Score conversion (PM2S) task~\cite{Liu2022performance}. To perform this task, we utilize the PM2S dataset compiled by Liu \emph{et al.}~\cite{Liu2022performance}, follow their train/valid/test split, and adopt the F1-score for evaluation. 
Since the input is not quantized in time, the pseudo-tokens of onset time and duration are obtained by the following pre-processing steps: 
1) determine the global tempo $T$ by assuming the length of one beat as the median value of all notes' duration; 
2) if $T\notin [40, 200]$, then $T$ is multiplied or divided by 2 to make it lie in the range; 3) quantize the note sequence to 16th notes by assuming a 4/4 time signature and constant tempo regardless of the actual performance content.\footnote{This does not imply that the actual time signature must be 4/4, but is simply a method to construct input tokens for the proposed model.}

\textbf{Downbeat note prediction (DbP).} Similar to BP, this task aims to identify downbeat notes. 
We again use the PM2S dataset with the same train/valid/test split as Liu \emph{et al.}~\cite{Liu2022performance} and use the F1-score metric.

\textbf{Chord root estimation (CR).} This task involves predicting the tatum-wise chord roots for symbolic music and serves as a crucial subtask in functional harmony analysis~\cite{chen2018functional, Lopez2021augmentednet, Karystinaios2023roman, Sailor2024rnbert}. We use the combined dataset compiled by AugmentedNet for this task and follow its train/valid/test split~\cite{Lopez2021augmentednet}. Following~\cite{Lopez2021augmentednet, Karystinaios2023roman, Sailor2024rnbert}, we also adopt the Chord Symbol Recall (CSR) metric, but at the 1/4 crotchet beat level (instead of 1/8), since both MidiBERT and the proposed M2BERT have a time resolution of 1/4 crotchet beats. This should only lead to a negligible difference, as chord changes rarely occur at 1/8 crotchet beats. 

\textbf{Local key estimation (LK)} estimates the local key for symbolic music and is also a crucial subtask of functional harmony analysis. 
We utilize the same dataset, experiment setting and evaluation metric as in the case of CR.

\textbf{Symbolic melody extraction (ME)} classifies each note into three classes:
vocal melody, instrumental melody, or accompaniment, which was also used for evaluation in MidiBERT~\cite{Chou2021midibert}. We use the POP909 dataset~\cite{Wang2020pop909} for this task and follow the same train/valid/test split as in MidiBERT. Note-level accuracy is adopted for evaluation.

\textbf{Symbolic velocity estimation (VE).}
Also introduced in MidiBERT, the objective of this task is to estimate the velocity of notes in terms six classes: \emph{pp}, \emph{p}, \emph{mp}, \emph{mf}, \emph{f}, and \emph{ff}. We use the same setting and evaluation metric as in ME. 

\textbf{Orchestral texture classification (OTC)} labels the textural layer of each track in symphony music in terms of \emph{melody}, \emph{rhythm}, and \emph{harmony}~\cite{Le2022corpus, Lin2024s3}. We utilize the dataset proposed by Le \emph{et al.} in~\cite{Le2022corpus}, which contains texture annotation of 24 symphony movements composed by Beethoven, Haydn, and Mozart. 18 of them are accompanied with digital score and have been used for the evaluation recently in~\cite{Chu2023orchestral}. 
We therefore follow the settings in~\cite{Chu2023orchestral}, including the 7-class formulation for the multi-label scenario (exclude the ``no label'' class), the same train/test split, and bar-level accuracy for evaluation. 

\textbf{Motif note identification (MNID).} Similar to ME, MNID classifies whether or not a note in a music belongs to a motif of that music. This task plays a critical role in music analysis by highlighting important notes that contribute to the motifs of a musical piece~\cite{Hsiao2023bps}. We adopt the BPS-motif dataset for this task~\cite{Hsiao2023bps}, and employ the note-level F1-score for evaluation. As a first attempt to this task, we divide this dataset into five folds and conduct cross-validation on it.

\textbf{Violin fingering prediction (VF)} predicts the fingering that a performer uses to perform a symbolic violin piece. This task could also be considered as a generation task, as the fingering choices are up to the performer themself. However, we still consider this task as an aspect of music understanding since it corresponds to the notion of \emph{playability}, a physical constraint over notes. We use the TNUA dataset~\cite{Jen2021positioning} for this task and use the same train/valid/test split as~\cite{Lin2024enhancing}. Following~\cite{Lin2024enhancing}, we treat violin fingering as a 240-class classification task (the combination of four string choices, five finger choices, and 12 hand position choices) and use note-level accuracy for evaluation.

\section{Experiment setup}\label{sec:experiment_setup}
We utilize the StableAdamW optimizer~\cite{Wortsman2023stable} and randomly corrupt 30\% of the note tokens for pre-training. The corruption hyperparameters $\textrm{r}_\textrm{pos}$, $\textrm{r}_\textrm{pit}$, and $\textrm{r}_\textrm{dur}$ are set to 4, 12, and 12, respectively. All other pre-training hyperparameters remain the same as MidiBERT, namely: learning rate of $2\times10^{-5}$, weight decay of 0.01, batch size of 12, 85\%/15\% partition of the training and validation set, and choose the best model checkpoint that maximizes the token reconstruction accuracy on the validation set.

For the pre-training dataset, we employ two settings. The first one, referred to as the ``Reduced'' dataset, follows MidiBERT~\cite{Chou2021midibert} by combining POP909~\cite{Wang2020pop909}, Pop1K7~\cite{Hsiao2021compound}, EMOPIA~\cite{Hung2021emopia}, Pianist8~\cite{Chou2021midibert}, and ASAP~\cite{Foscarin2020asap}, while excluding all non-4/4 time signature music from the training data.\footnote{We use this setting to have a fair comparison with MidiBERT. However, as discussed in Section~\ref{subsec:tokenization}, the proposed model can still process non-4/4 music pieces by rescaling the bars, which are frequently seen in tasks that involve classical music, such as CR, LK, and MNID.} This leads to a dataset containing 4.89M notes. For this setting, we train the model for 150 epochs, which takes around 18 hours on an NVIDIA RTX 6000 Ada GPU.\footnote{The original MidiBERT trains the model for 500 epochs, but we found that for the proposed M2BERT, the validation loss usually reaches the lowest point between 100 and 150 epochs. Therefore, we terminate the training at the end of 150 epochs.} The second setting, referred to as the ``Full'' dataset, contains the Reduced dataset plus the \emph{lmd\_matched} set of the Lakh MIDI dataset~\cite{Raffel2016learning}, totaling 350.32M notes. Because of the limitation of our computational resources, we only train the model for 25 epochs when using the Full dataset, regardless of the decreasing validation loss.
In the experiments, we employ the Reduced setting in all the ablation studies for 
a fair comparison with MidiBERT, 
while only employing the second setting for our best configuration. 


\begin{table*}[ht]
\fontsize{10pt}{10pt}\selectfont
\setlength{\tabcolsep}{2.7pt}
\centering
\begin{tabular}{c|c|c|c|cc|cc|cc|c|c|c|c|c}
  \toprule
  & & & SGC & BP & DbP & CR & LK & ME & VE & OTC & PS & ER & VF & MNID\\
  Model & Objectives & Dataset & Acc. & F1 & F1 & CSR & CSR & Acc. & Acc. & Acc. & Acc. & Acc. & Acc. & F1\\
  \midrule
  MidiBERT & MLM & Reduced & .397 & .843 & .706 & .795 & .761 & .964 & .518 & .694 & .708 & .638  & \textbf{\textit{.551}} & .700\\
  M2BERT & MLM & Reduced & .403 & .867 & .749 & .825 & .796 & \textit{.975} & .524 & \textit{.695} & \textit{.740} & .641 & \textit{.509} & .708\\
  M2BERT & $\text{RC}_{\infty}$ & Reduced & .397 & .864 & \textit{.768} & .838 & .780 & \textit{.978} & \textbf{\textit{.536}} & \textit{.703} & \textit{.740} & .666 & \textit{.538} & .712\\
  M2BERT & $\text{RC}_{4, 12, 12}$ & Reduced & .404 & .867 & \textit{.779} & .838 & .791 & \textit{.983} & .534 & \textit{.707} & \textit{.740} & .658 & \textit{.511} & .719\\
  M2BERT & $\text{RC}_{4, 12, 12}$+Pianoroll & Reduced & .405 & .867 & \textit{.768} & .847 & .804 & \textit{.982} & .532 & \textbf{\textit{.725}} & \textit{.742} & .667 & \textit{.536} & .718\\
  M2BERT & $\text{RC}_{4, 12, 12}$+Pianoroll & Full & \textbf{.433} & \textbf{.886} & \textbf{\textit{.839}} & \textbf{\textit{.855}} & \textbf{.811} & \textbf{\textit{.986}} & .534 & \textit{.720} & \textbf{\textit{.746}} & \textbf{.671} & \textit{.539} & \textbf{.745}\\
  \midrule
  SOTA & NA & NA & - & .945 & .757 & .849 & .829 & .974$\ast$ & .536 & .695 & .736$\dagger$ & .719$\dagger$ & .489 & -\\
  \bottomrule
\end{tabular}
\caption{Evaluation results on the proposed SMC benchmark. If possible, we also show the state-of-the-art (SOTA) performance on the tasks reported by previous work. $\ast$ in the SOTA row means that it is reported under a simpler (2-class instead of 3-class) problem definition~\cite{Zhao2023multi}; $\dagger$ denotes that the results are reported with a different dataset partition~\cite{Shen2023more}. Results that achieve or surpass the SOTA are shown in italic, and the best results among pre-trained models are shown in bold. The SOTAs are obtained from~\cite{Sailor2024rnbert, Karystinaios2023roman, Zhao2023multi, Shen2023more, Chu2023orchestral, Lin2024enhancing}. For BP and DbP, we reproduce the evaluation of~\cite{Liu2022performance} and achieve a significantly better results than what they report (0.862 and 0.698, respectively), so we show the reproduction results.}\label{tab:downstream_results}
\end{table*}

\begin{table*}[ht]
\fontsize{10pt}{10pt}\selectfont
\setlength{\tabcolsep}{3.5pt}
\centering
\begin{tabular}{c|c|c|c|cc|cc|cc|c|c|c|c|c}
  \toprule
  Model & Objectives & Length & SGC & BP & DbP & CR & LK & ME & VE & OTC & PS & ER & VF & MNID\\
  \midrule
  M2BERT & $\text{RC}_{4, 12, 12}$+Pianoroll & 512 & .403 & .870 & .765 & .808 & .779 & .976 & .533 & .693 & .733 & .647 & .493 & \textbf{.725} \\
  M2BERT & $\text{RC}_{4, 12, 12}$+Pianoroll & 1024 & \textbf{.405} & .867 & .768 & .847 & .804 & .982 & .532 & \textbf{.725} & .742 & \textbf{.667} & \textbf{.536} & .718\\
  M2BERT & $\text{RC}_{4, 12, 12}$+Pianoroll & 2048 & .397 & \textbf{.887} & \textbf{.773} & \textbf{.862} & \textbf{.805} & \textbf{.985} & \textbf{.538} & .710 & \textbf{.746} & .653 & .527 & .719\\
  \bottomrule
\end{tabular}
\caption{The ablation study results on the choice of maximum sequence lengths (the ``Length'' column). }\label{tab:ablation}
\end{table*}

Our comparative study incorporates two models (i.e., MidiBERT and M2BERT), two datasets (i.e., Reduced and Full), and three training objectives:
1) the original MLM objective; 2) the proposed corrupted note reconstruction objective (denoted as $\text{RC}_{(\textrm{r}_\textrm{pos}, \textrm{r}_\textrm{pit}, \textrm{r}_\textrm{dur})}$, parametrized by the corruption ranges $(\textrm{r}_\textrm{pos}, \textrm{r}_\textrm{pit}, \textrm{r}_\textrm{dur})$ labeled in the subscripts); 3) the pianoroll prediction objective (denoted as Pianoroll).
It should be noted that $\text{RC}_{\infty}$ denotes note corruption by random sampling on all possible tokens. The source code of the proposed model and additional ablation study results are available at \url{https://github.com/york135/M2BERT}.

\section{Results}\label{sec:results}

Table~\ref{tab:downstream_results} presents the experiment results over all the settings on the 12 downstream tasks. For reference, we show the state-of-the-art (SOTA) performance of the tasks in the last row, if available.
First, we observe that all the settings of the proposed M2BERT model outperforms MidiBERT for almost all the downstream tasks; this demonstrates the advantages of the ModernBERT architecture.
Second, random sampling (i.e. $\text{RC}_{\infty}$) does outperform MLM, possibly due to the removal of the \texttt{[MASK]} token. However, the proposed $\text{RC}_{4, 12, 12}$ further outperforms $\text{RC}_{\infty}$; this shows the effectiveness of limiting the range of random corruption. Finally, incorporating the Full dataset for training further improves overall performance.

Further delving into the results on each downstream task, we found that the individual proposed objectives do not consistently benefit all downstream tasks. Comparing MLM and $\text{RC}_{4, 12, 12}$+Pianoroll (under M2BERT), we found that the improvements of $\text{RC}_{4, 12, 12}$+Pianoroll mainly lies in DbP, CR, ME, OTC, and ER. Remarkably, the ME accuracy is improved from 0.975 to 0.982, which represents a relative error reduction of 28\%. However, in other tasks, the improvements are marginal or even zero. We suspect that the proposed objectives tend to represent the musical domain knowledge related more on pitches, intervals, and chords, and therefore benefit more on these tasks (note that DbP is also benefited, possibly because it is somewhat related to chord changes). On the other hand, for BP, pitch and pitch interval are less important (as discussed in~\cite{Liu2022performance}) and are therefore not benefited.

Another interesting finding lies in the result of VF, the only task in which MidiBERT outperforms the proposed M2BERT. 
While this suggests that M2BERT is not universally better than MidiBERT, we should also take into account that 1) the fine-tuning dataset for VF is smaller than those for other tasks; 2) VF is different from the other tasks in the sense that VF focuses more on the performer's empirical preferences rather than music theory. These factors, rather than the model itself, might be responsible for this trend. Nevertheless, we can still observe that the proposed note correction and pianoroll prediction objectives do improve the performance of VF for M2BERT.

Furthermore, our best model, M2BERT with the $\text{RC}_{4, 12, 12}$+Pianoroll objectives, pre-trained with the full dataset, outperforms the SOTAs in DbP, CR, LK, ME, OTC, PS, and VF (six of the ten tasks that have a comparable SOTA). Considering that 1) we do not employ multitask learning or data augmentation, both are frequently used in SOTA works~\cite{Liu2022performance, Sailor2024rnbert, Karystinaios2023roman}; 2) we do not adopt task-specific model design or hyperparameter search; and 3) we also do not adopt task-specific input representation; these results are promising, showing the effectiveness of using a pre-trained model for symbolic music classification. 

Finally, we conduct an ablation study on the choice of different maximum sequence lengths for the proposed M2BERT model. The results are shown in Table~\ref{tab:ablation}.
We observe that for most of the downstream tasks, using a longer token length leads to better performance. Specifically, the two tasks which are benefited most by using a long token length are CR, which is improved by 5.4 percentage points, and LK, which is improved by 2.6 percentage points, when increasing the sequence length from 512 to 2048, at the cost of the quadratic time complexity introduced by the Transformer architecture. Considering such a trade-off, we eventually choose the sequence length of 1024 in all other experiments (see Table~\ref{tab:downstream_results}).

\section{Conclusion}\label{sec:conclusion}
With a systematic evaluation on a benchmark incorporating 12 downstream tasks, we have demonstrated the effectiveness of using token denoising and pianoroll prediction to enhance the pre-training of a BERT-like model for symbolic music understanding. This result underscores the important insight that symbolic music pre-training should focus on optimizing meaningful musical information, and, on the other hand, reveals the limitations of applying traditional NLP methods that optimize token sequences to symbolic music data. We also show the effectiveness of using an advanced model architecture and increasing the scale of data for pre-training.
Additionally, considering that our symbolic pre-trained model was only trained on at most 350M notes—whereas text pre-trained models can leverage trillions of text tokens~\cite{Touvron2023llama, Dubey2024llama}—there is significant potential for scaling up our approach given more data~\cite{Hestness2017deep}.
It is also important to clarify that, to focus the discussions on model pre-training strategies, our current benchmark consists solely of classification tasks that allow for direct linear probing. 
Therefore, to further advance symbolic music foundation models in music understanding, our future directions include developing methods to optimize other types of musical information (e.g., rhythm), addressing more integrative symbolic music understanding tasks such as the entire functional harmony recognition task and the performance MIDI-to-score conversion task, and identifying pathways for further scaling up our model.

\section{Ethics Statement}\label{sec:ethics_statement}
In this work, all the datasets used in model training and evaluation are publicly available and can be downloaded without submitting any data access application form. While this strict policy improves the reproducibility, it also makes the evaluation biased toward high-resource tasks and music genres that have a larger amount of publicly available data. For example, for classical music, we mainly focus on western classical music instead of classical music from other regions. Future efforts could be made to mitigate this issue by adding more diverse datasets and downstream tasks to the evaluation benchmark.

\section{Acknowledgments}
This work is supported in part by National Science and Technology Council under Grant NSTC 113-2221-E-001-013, the Academia Sinica Grand Challenge (GCS) Program under Grant AS–GCS–112–M07, and the Postdoctoral Scholar Program of Academia Sinica under Grant AS-PD-1141-M15-2.

\bibliography{ISMIRtemplate}

%
%
%
%

\end{document}